\documentclass[osajnl,twocolumn,showpacs,superscriptaddress,10pt]{revtex4-1} %% use 11pt for Applied Optics
\usepackage{epsf,epsfig}
\usepackage[psamsfonts]{amssymb}

\usepackage{amsmath}
\usepackage{bm}
\usepackage{graphicx}
\usepackage{graphicx,epstopdf}
\usepackage{caption}
\usepackage{subcaption}
\newcommand{\bfsfG}{\mbox{\sffamily\bfseries{G}}}

\begin{document}

\title{Entanglement of two two-levels atoms mediated by an optical black hole}

\author{Mahnaz Tavakoli}
\affiliation{ Department of Physics, Faculty of Basic Sciences, Shahrekord University, P.O. Box 115, Shahrekord 88186-34141, Iran.}

\author{Ehsan Amooghorban}\email{Corresponding author: Ehsan.amooghorban@sci.sku.ac.ir}
\affiliation{ Department of Physics, Faculty of Basic Sciences, Shahrekord University, P.O. Box 115, Shahrekord 88186-34141, Iran.}
\affiliation{ Photonic Research Group, Shahrekord University, P.O. Box 115, Shahrekord 88186-34141, Iran.}

\begin{abstract}
We consider the dynamics of a system consisting of two two-level atoms interacting with
the electromagnetic field near an optical black hole.
We obtain the reduced density operator of the two-atom system in the weak coupling regime for the case that one atom is in the excited state and the other in the ground state. The time evolution of the negativity between the atoms is discussed for two non-resonance and resonance cases.
In both cases, we show that the two atoms can become entangled due to the indirect interaction mediated through the optical black hole.
\\
\\
\end{abstract}

\maketitle

\section{Introduction}
One of the most striking predictions of the general relativity is undoubtedly the possibility of existence of black holes (BHs). Various astrophysical observations have confirmed the existence of BHs with almost certainty~\cite{Eckart1997,Abbott2016a,Abbott2016b}. The BHs are classically described as massive objects with such a strong gravitational field that no signals, not even light escape from a region so-called event horizon.
Hawking discovered that BHs are not completely black and emit particles in the form of thermal radiation due to quantum effects~\cite{Hawking1974}.
Despite recent technological developments, such phenomenon would be seem to be impossible to observe directly using astronomical tools due to the low Hawking temperatures associated with gravitational black holes.
%
%On other hand, the analysis of BHs and the study of their physics is an active and important area of research in theoretical physics.
Therefore, many attempts have been made to circumvent such problems and to mimic certain aspects of these celestial objects in analogue systems~\cite{Unruh1981,Barcelo2005}, such as the
Bose-Einstein condensate~\cite{Garay2000,Barcelo2001}, moving dielectrics~\cite{Lorenci2003}, optical fiber~\cite{Philbin2008},
superconducting transmission line~\cite{Nation2009} and more recently magnetization dynamics~\cite{Molina2017}.
%Although, analogue models of general relativity have long ago been considered in the literature~\cite{Gordon1923}.

Among these, the recent developments in transformation optics have provided the analog of the bending of light in empty curved space-time caused by gravity field with the aid of metamaterials. Due to the formal invariance of Maxwell's equations under transformation optics and as well the analogy between the Maxwell's equations in the presence of anisotropic and inhomogeneous media and free-space Maxwell's equations in curved space-time~\cite{Eddington1920,Gordon1923,Plebanski1960,Leonhardt2006},
%transformation optics provide us an excellent mathematic tool for designing a vast range of the optical analogues of cosmic phenomena.
%Metamaterials are artificial materials composed of subwavelength structures to provide exotic effective macroscopic electromagnetic behavior not generally found in nature. They offer considerable flexibility in controlling light-matter interactions and light propagation in unusual ways.
%
an isotropic optical BH (OBH) was suggested to reproduce the behavior of BH in laboratory and their consequences investigated~\cite{Genov2009}. In another approach from Hamiltonian optics, Narimanov and Kildishev have proposed a broadband absorber device that acts like an effective OBH with the event horizon radius is determined by the matter's boundary~\cite{Narimanov2009}.
The device was composed of two parts: a core with the constant electric permittivity and an outer shell with an inhomogeneous and isotropic electric permittivity. The outer shell can appropriately guide the electromagnetic waves to the core and then the incident waves absorb or harvest by the core completely.
%
%In 2010, based on this model, an OBH has been simulated by replacing the shell with a multilayered structure with equal thickness~\cite{Lu2010}.
Attempts to realize the OBH idea have been made numerically by full-wave simulations~\cite{Kildishev2010,Liu2010,Argyropoulos2010,Lu2010}, and experimentally by using nonresonant and resonant metamaterial structures~\cite{Cheng2010,Zhou2011,Yang2012} and three-dimensional woodpile photonic crystals structure~\cite{Yin2013} in the microwave frequency. The results validated their broadband performance and demonstrated that these designed structures can effectively absorb the incident waves from all directions. The capability of such devices in capturing and absorbing the broadband and omnidirectional electromagnetic wave may find potential applications in solar energy harvesting, radiation detector, and optoelectronics~\cite{Landy2008,Atwater2010,Schuller2010}.

%This system forms an excellent theoretical laboratory where many of the unknown effects that quantum gravity could exert on black-hole evaporation can be  modelled.

%This model also admits the possibility of experimental testing, although this is an extremely slim possibility.

So far, all attempts in the context of OBHs based on Narimanov model are limited to
%Until now, the concept of the optical BH based on Narimanov model has been applied to
control and trapping the electromagnetic waves in classical framework around a cylinder or sphere core with engineering
materials, similar to that around BHs in general relativity.
However, there is another interesting possibility by treating light as a stream of photons rather than electromagnetic waves when the light interacting with OBHs. The inspiration for this work comes from our earlier study of the entanglement dynamic and radiative properties of an atomic system near an invisibility clocking device~\cite{Morshed2016,Amooghorban2017}.
The fluctuating electromagnetic field induces noise currents within material media. These noise currents act as a source for the quantized electromagnetic field. Thus, by investigating the interaction of a atomic system with the quantized electromagnetic fields we can examine the effect of the OBHs on internal properties of the atomic system. In this sense, the atomic system can be treated as an open quantum system coupled to the environment, i.e., with the electromagnetic field in presence of material media, that leads to dissipation and decoherence.
As a consequence, the quantum entanglement may disappear and even enhance in certain circumstances.
%Quantum entanglement is an important physical phenomenon in which the quantum states of several objects cannot be described independently.
%we will also try to see what effects the uncertain entangled state will have on the degradation of entanglement in our scheme due to the presence of an arbitrary state parameter. which are fundamental building blocks of
%Thus, the quantum entanglement, which is a quite amazing physical phenomenon, can be studied by the dynamical evolution of the atomic system.

With the above background and taking into account that the entanglement play a key role in gravitational BH, in this paper, we examine the influences of the OBHs in terms of the entanglement created in an atomic system, in order to study the role of the OBH effects from a quantum perspective.
%This allows us to gain a new understanding of the optical BH effects from a quantum perspective.
The atomic system we are going to study consists of two identical and mutually independent two-level atoms with one initially in its excited state and the other in its ground state and weakly interact with the fluctuating quantized electromagnetic fields in vacuum outside an OBH.
In the absence of the OBH, quantum entanglement arises from the spontaneous emission process and the
mutual dipole-dipole coupling of the atoms~\cite{Tanas2003,Tanas2004}. In the presence of the OBH, two noninteracting quantum systems can become entangled due to the photon exchange process mediated through this OBH, of course, if the photon is not absorbed by the OBH.
We therefore expects that the dynamical behavior of entanglement for the atomic system becomes drastically different from what would be experienced in free space.

This paper is organized as follows. In Sec.~\ref{Sec:The basic relations}, we introduce the model and give a review of the general expressions needed to describe the system of two two-level atoms coupled with quantized electromagnetic field near an OBH. This OBH which defined with continuous material parameter can be readily implemented by a large number of thin layers with homogeneous material parameters in a stepwise manner.  In so doing, we are not only able to calculate the Green's tensor of the system using the formalism developed by~\cite{Tai1994}, but can also serve as a new approach to realize the OBH by concentric layered structures instead of using the metamaterial with subwavelength resonant inclusions.
%The dynamical evolution of entanglement between two polarizable two-level atoms in weak inter- action with electromagnetic vacuum fluctuations is investigated.
We then study the time evolution of the two atoms that initially share a single excitation, and the collective behavior of the atoms is demonstrated in Markovian regimes.
%In Sec.~\ref{Sec:The basic relations}, with the suppose of weakly interacting with the EM field, the
%the densitymatrix equations of motion for the system that consists of the atoms and the resonant part of the electromagnetic field are derived.
%
In Sec.~\ref{Sec:Entanglement}, the dynamical evolution of entanglement between the two atoms, measured by negativity, is discussed both in the presence of the OBH, as well as in the absence of it, and the influences of material absorption, resonant and off-resonant coupling of the atoms
to the electromagnetic field are analyzed.
Finally, a summary of the results are given in Sec.~\ref{Sec:conclusion}. Details on the Green's tensor of the system can be found in Appendix~\ref{App:Green tensor}.

\begin{figure}[t]
\includegraphics[width=1\columnwidth]{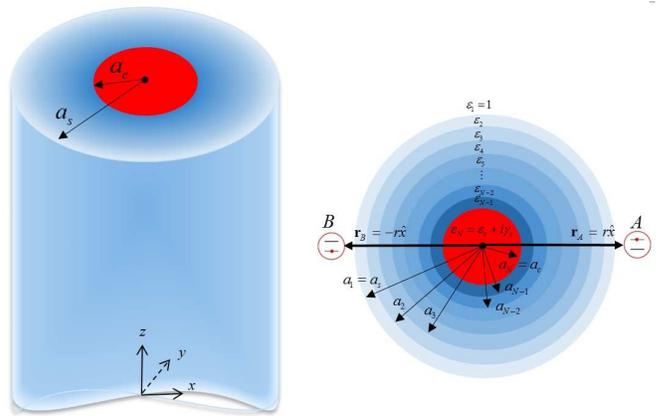}
\caption{Pictorial illustration of the system. The OBH is composed of a circular shell and a lossy dielectric core, and placed in free space. $a_c$ is the radius of the inner absorbing core, $a_s$ is the radius of the outer shell which denotes the event horizon. The outer shell is implemented in a concentric layers structure with $N-2$ layers whose the material parameter in each layer is given by Eq.~(\ref{piecewise-constant permittivity function}). $\varepsilon_1$ is the permittivity of the background medium and $\varepsilon_N$ is the complex permittivity of the absorbing core. Two two-level distant atoms are placed at positions ${\bf r}_A$ and ${\bf r}_B$ from the center of the OBH.}\label{Fig:1}
\end{figure}
%

%%%%%%%%%%%%%%%%%%%%%%%%%%%%%%%%%%%%%%%%%%%%%%%%%%%%%%%%%
%%%%%%%%%%%%%%%%%%%%%%%%%%%%%%%%%%%%%%%%%%%%%%%%%%%%%%%%
\section{The basic relations}\label{Sec:The basic relations}
%%%%%%%%%%%%%%%%%%%%%%%%%%%%%%%%%%%%%%%%%%%%%%%%%%%%%%%%%%
%With the preceding discussion,

In this section, we start with a brief description of the basic features of an OBH based on Narimanov model. This OBH consisting of a lossy inner core and an outer shell with spatially varying values of permittivity. Such material properties can be typically realized by some kind of resonance-like structures which suffer from inherently loss and dispersion. Therefore, to have physically values for the dielectric permittivity of the shell, we must treat $\varepsilon$ as a frequency dependent function in this region.
Suppose that the permittivity function has the Lorentzian type dispersion characteristics in inhomogeneous region and the device is placed in free space, hence, the permittivity of the surrounding is unit. The permittivity profiles of the OBH can be described as follows:
\begin{eqnarray}\label{material parameters of optical BH}
\varepsilon (r) = \left\{ {\begin{array}{*{20}{c}}
{1,\quad \quad }&{r > {a_s}},\\
{{{(\frac{a_s}{r})}^2}\varepsilon_L,\quad }&{{a_c} < r < {a_s}},\\
{{\varepsilon _c} + i\gamma_c, \quad \quad  }&{r < {a_c},\quad \quad }
\end{array}} \right.
\end{eqnarray}
where $a_{c} $ and $a_{s} $ denote the radii of inner core and outer shell of the OBH, respectively, and $ \varepsilon_L=\left( {1 + \frac{{\omega _p^2}}{{\omega _0^2 - {\omega ^2} - i\gamma \omega }}} \right) $ is the Lorenzian form of dispersion of the outer shell, wherein ${{\omega }_{p}}$ and ${{\omega }_{0}}$ are, respectively, the plasma and resonance frequency and $\gamma $ is the absorbtion coefficient. Besides, the radius of the core, ${a_c}$, and the radius of the outer shell satisfy the relation: ${a_c} = {a_s}\sqrt \frac{1}{\varepsilon _c}$.
Here, we use the suggested parameters in~\cite{Lu2010} for an OBH with the inner core radius $4\pi c/\omega_0$, the outer shell radius $8\pi c/\omega_0$, and the permittivity of the inner absorbing core $4+0.33i$.
%The quantum system is inevitably influenced by the surrounding environment under the usual conditions. The interaction between environment and quantum system can cause decoherence and lead to entanglement decay and death.
%
%Let us consider an atomic system formed by

We consider a system consisting of two equal two-level atoms $A$ and $B$ with dipole moments $ {\bf d}_A={\bf d}_B={\bf d} $, which are assumed to be directed along the OBH axis, i.e., $ {\bf d} =d{\bf z}$. The atoms with two stationary states $|l_{A,B}\rangle$ and $|u_{A,B}\rangle$ are symmetrically located at positions ${\bf r}_A = -{\bf r}_B={ r}\hat{{\bf x}}$ on the $x $ axis close to the OBH. The spacing energies of the two atoms are denoted by $\hbar\omega_A$.
The atoms are in vacuum outside the OBH and interact with the quantum vacuum electromagnetic field via their transition dipole moments. This provide indirect interaction mediated through the OBH between the atoms.

To quantum mechanically describe the aforementioned system, we follow the canonical quantization of the electromagnetic field in the presence of absorptive and dispersive dielectric medium. Based on this approach, the medium is directly introduced to the quantization process by modeling it through a continuum of reservoir oscillator field to account for the dissipation and polarizability characters of the matrical medium.
With a freedom of choice, we start with an appropriate Lagrangian to describe two identical two-level atoms interacting with fluctuating electromagnetic
fields in the presence of the medium.
%
%For further details concerning
For a thorough discussion of the total Lagrangian of the system, the interested reader is referred to~\cite{Morshed2016,Amooghorban2017,Huttner 1992,Jeffers1996,Suttorp2004,Amooshahi 2009,Kheirandish 2010,Philbin 2010,Amooghorban 2011,Amooghorban arXiv}.
We use the total Lagrangian and define the canonical conjugate momentums of the system, in such a case, the total Hamiltonian of the coupled system that governs the evolution of the system is obtained in the electric-dipole and rotating wave approximations as~\cite{Philbin 2010,Kheirandish 2011}
\begin{eqnarray}\label{total Hamiltonian}
\hat H &= &\int {{d^3}r\int_0^\infty  {d\omega \, \hbar\omega {{{\bf{\hat f}}}{^\dag} }} } \, ( {{\bf{r}},\omega } )\cdot{\bf{\hat f}} ( {{\bf{r}},\omega } ) + \sum\limits_{j = A,B}  \hbar {\omega_j }{{\hat \sigma }^\dag }_j {{\hat \sigma }_j }\nonumber \\
&& -\sum_{j = A,B} \Big( {{{\hat \sigma }^\dag }_j {{\bf{d}}_j }\cdot\int_0^\infty  {d\omega \hat {{\bf{E}}}\left( {{\bf{r}}_j,\omega } \right)}  + h.c.} \Big),
\end{eqnarray}
where $ {\hat \sigma_j } = | {{l_j }} \rangle \langle {{u_j }} | $
and $ {\hat \sigma}_j ^\dag  = \left| {{u_j }} \right\rangle \left\langle {{l_j }} \right| $ are, as usual, the lowering and raising Pauli operators of $j$-th atom, $ {\bf d}_j  = \left\langle {{u_j }} \right|{\rm{\bf d}_j}\left| {{l_j }} \right\rangle $ is the matrix
element of the dipole moment operator ${\bf d}_j$ of $j$-th atom, and  $ {\bf{\hat f}}^\dag({\bf{r}},\omega ) $ and $ {\bf{\hat f}}({\bf{r}},\omega ) $ denote the bosonic creation and
annihilation operators which play the roll of the collective excitations of the electromagnetic field and the medium.
The transition to the quantum regime can be done in a standard fashion by imposing the commutation relations between the variables and their conjugates. It was shown that these commutation relations eventually lead to the usual commutation relations of bosonic operators~\cite{Morshed2016,Kheirandish 2010,Amooghorban arXiv}:
\begin{eqnarray}\label{commutation relations f}
&& \left[ \hat{f}_{ j}\left( \bf{r},\omega  \right), \hat{{f}}_{ {j}'}^{\dagger }\left( \bf{{r}'},{\omega }' \right) \right]={{\delta }_{j{j}'}}\delta  ( \bf{r}-\bf{{r}'}  )\delta  ( \omega -{\omega }'  ),\nonumber \\
&& \left[ \hat{f}_{ j} ( {\bf r},\omega   ),\hat{f}_{j}' ( {\bf r'}, \omega'  ) \right]=0.
\end{eqnarray}
The positive frequency part of the electric field operator is expressed in term of $ \hat{\bf{f}}$  as
\begin{eqnarray}\label{E+}
{\bf E}^{(+)}\left(\mathbf{r},\omega  \right)&=& i\frac{\omega^2}{c^2} \int{{{\mbox d}^{3}}}r'\, \sqrt{\frac{\hbar\, {\rm Im}[\varepsilon({\bf r}',\omega)]}{\pi\varepsilon_0}}\, \bfsfG  ({ \bf r},{\bf r'},\omega )\nonumber\\
&&\cdot \hat{{\bf f}} ({\bf r'},\omega   ),
\end{eqnarray}
where $\bfsfG ( \mathbf{r},\mathbf{r'},\omega )$ is the classical Green's tensor satisfying the Helmholtz equation
\begin{eqnarray}\label{Helmholtz equation}
&& \nabla \times \nabla \times \,\bfsfG\left( \bf{r},\bf{r'},\omega  \right) \nonumber\\
&& -\frac{\omega^2}{c^2}{{ \varepsilon }}\left(\bf{r},\omega  \right)\bfsfG\left( \bf{r},\mathbf{r'},\omega  \right)=\delta \left( \bf{r}-\bf{r'} \right)\bar{\bar{\bf I}},
\end{eqnarray}
together with the boundary condition $\bfsfG ({\bf r}',{\bf r},\omega)\rightarrow 0$ for $|{\bf r}-{\bf r}'| \rightarrow \infty$. Owing to
the symmetry of the dielectric function, the Green's tensor is reciprocal, $\bfsfG ({\bf r},{\bf r}',\omega)=\bfsfG^T ({\bf r}',{\bf r},\omega)$, it is analytic in the upper half of the complex $\omega$ plane, and like every causal response function it obeys the Schwarz reflection principle $\bfsfG^* ({\bf r},{\bf r}',\omega)=\bfsfG ({\bf r},{\bf r}',-\omega^*)$. It contains all the information about the geometry and and topology
of the system.

For single quantum excitation, the time-dependent state vector of the whole system can be written as
\begin{eqnarray}\label{time-dependent state vector of the system}
&&| \psi ( t ) \rangle =\sum\limits_{j  = A,B} {C_{U,j}}\left( t \right){e^{ - i{(\omega_j-{\tilde \omega }})t}}\left| U_j \right\rangle \left| \{0\} \right\rangle
\nonumber \\
&&+  {\int {{d^3}r} } \int_0^\infty
{d\omega \,\,{e^{ - i(\omega-\tilde{\omega} )t}}{{ C}_{L }}} \left(
{{\bf{r}},\omega ,t} \right)  \left| L   \right\rangle \left|{{{\bf{1}} }\left(
{{\bf{r}},\omega } \right)} \right\rangle,
\end{eqnarray}
where $\bar{\omega }=\sum\limits_{j}{{{\omega }_j}}/2$, and the first element of the state vector indicates the state of the atoms and the second
element that of the field. Here, the state vector $| U_j \rangle$ denotes the $j$th atom is in the
excited state and the other atom is in the ground state, i.e., $| U_A \rangle=| u_A,l_B \rangle$ and $| U_B \rangle=| l_A,u_B \rangle$, the state vector $| L \rangle=|l_A,l_B \rangle$ refers to both atom are in the lower state, $| \{0\} \rangle$ is the vacuum state of the field, $| \{{\bf 1}({{\bf r},\omega })\} \rangle$ is the excited state of the field with the field is in a single-quantum Fock state, and ${{C_{U,j }}(t)} $ and $ {C_{L}}(\rm{\bf{\bf r}},\omega ,t) $ are, respectively, the respective probability amplitudes of the excited and ground states of the system.

In the Schr\"{o}dinger equation picture, the evolution of the state vector of the system  at any time $t > 0$ obeys
the Schr\"{o}dinger equation $i\hbar\, {\partial}/{\partial t} | \psi ( t ) \rangle={\cal H} | \psi ( t ) \rangle$. By inserting Eq.~(\ref{time-dependent state vector of the system}) into the Schr\"{o}dinger equation coupled motion equations for the expansion coefficients $C_{U,A } $ and $C_{U,B } $ are obtained. The details of these calculations can be found in~\cite{Amooghorban2017}.
It is convenient to introduce the new variables, ${{C}_{\pm }}\left( t \right)=\left[ {{C}_{{{u}_{A}}}}\left( t \right)\pm {{C}_{{{u}_{B}}}}\left( t \right) \right]/\sqrt{2}$,
which are the probability amplitudes of finding the atomic subsystem in the collective symmetric and antisymmetric states $\left| \pm  \right\rangle =(\left| {{u}_{A}},{{l}_{B}} \right\rangle \pm \left| {l}_{A},{{u}_{B}} \right\rangle )/\sqrt{2}$, to decouple the motion equations from each other. To simplify our calculations, let us restrict our attention the case when atom-field system
is coupled weakly.
This allow us to apply the Markov approximation and obtain analytical expressions for the symmetric and antisymmetric probability amplitudes ${{C}_{\pm }}$ as follows:
\begin{eqnarray}\label{time evolution C+-}
{C_ \pm }(t) = {e^{({{ - {\Gamma _ \pm }} \mathord{\left/
{\vphantom {{ - {\Gamma _ \pm }} 2}} \right.
\kern-\nulldelimiterspace} 2} + i{\delta _ \pm })t}}{C_ \pm }(0),
\end{eqnarray}
where ${{\Gamma }^{\pm }}={{\Gamma }}\pm {{\Gamma }_{AB}}$ and ${{\delta }^{\pm }}={{\delta }}\pm {{\delta }_{AB}}$ are, respectively, the decay rates and level shifts of the symmetric and antisymmetric states.
Here, the Lamb shift, $\delta={{\delta }_{jj}}$, is due to the atom electromagnetic self-interaction (radiation reaction) in the presence of the OBH, while, the level shift, ${{\delta }_{j\neq j'}}$, induced by the dipole-dipole coupling. Their explicit form are given by
\begin{eqnarray}\label{deita ij}
&& {{\delta }_{jj'}}=\frac{1}{\hbar \pi {{\varepsilon }_{0}}}P\int_{0}^{\infty }{d\omega' }\,\frac{{{\omega' }^{2}}}{{{c}^{2}}}\frac{{{\mathbf{d}}_{j}}\cdot \operatorname{Im} \left[ \bfsfG \left( {{\mathbf{r}}_{j}},{{\mathbf{r}}_{j'}},\omega_A  \right) \right] \cdot {{\mathbf{d}}_{j'}}}{(\omega' -{{\omega_A }})},\,\,\,\,\,\,\,\,\,
\end{eqnarray}
with $P$ denoting the principal value. The single-atom decay rate, $\Gamma= {{\Gamma }_{jj}}$, and the collective damping rate, ${{\Gamma }_{j\neq j'}}$, in Eq.~(\ref{time evolution C+-}) are given by
\begin{eqnarray}\label{Gamma ij}
{{\Gamma }_{jj'}}=\frac{2\,\omega_A^{2}}{\hbar {{\varepsilon}_{0}}{{c}^{2}}}{{\mathbf{d}}_{j}}\cdot \operatorname{Im} \left[ \bfsfG \left( {{\bf r}_j},{{\bf{r}}_{j'}},\omega_A  \right) \right] \cdot {{\mathbf{d}}_{j'}}.
\end{eqnarray}
The above expressions show the effect of the OBH on radiation properties of the atoms via the Green tensor of the system evaluated at
frequency $\omega$ and at positions ${\bf r}_j$ and ${\bf r}_{j'}$. Now, all that is needed is knowledge about the Green's tensor of the system explicitly. This is indeed a very complex problem, whose complexity arises from the difficulties in finding the appropriate cylindrical wave vector functions for inhomogeneous
region of the OBH. Instead, we model the aforementioned artificial BH by a multilayered cylindrical structure with equal thickness and, thereby, reduce the problem to the calculation of the electromagnetic Green's tensor of a dielectric multilayer cylinder.
This is because of the fact that we know how to calculate the Green's tensor of such structure.
%This method also used in the design of some invisibility cloaking devices~\cite{}
%

In the following, we discretize the inhomogeneous
region of the OBH into $N-2$ concentric shells of dielectrics with piecewise-constant permittivity function, as schematically illustrated in Fig.~\ref{Fig:1}. According to Eq.~(\ref{material parameters of optical BH}), we assume that the dielectric function of the $m$-th layer is given by
\begin{equation}\label{piecewise-constant permittivity function}
\varepsilon_m (\omega ) = \frac{{{a_1}^2}}{{{a_m}^2}}\varepsilon_L .
\end{equation}
where $ a_m (m=2,3,...,N-1)$ is the inner spherical interface of the $m$-th layer and $a_1=a_s$ is the radius of the outer shell.
Here, we use a 10-layer structure to implement such OBH as the thickness of each layer is $0.4\pi c/\omega_0$.

In the case of a dielectric multilayer circular cylinder, the electromagnetic Green's tensor
has been developed by Tai~\cite{Tai1994} and reconsidered in~\cite{Li2000}.
We briefly presented in appendix~\ref{App:Green tensor} the details involved in the derivation of the required Green's tensor, as well as of the vector eigenfunctions used to represent the free space and scattered contributions.
%
%In appendix~\ref{App:Green tensor}, we has given the bare essentials needed for calculating the electromagnetic Green's tensor of the system.
Due to the fact that in our case the dipole moments are directed along the $z$ axis, we only need the diagonal $z$ component of the Green's tensor. Considering the above points and the fact that the atoms are placed in free space outside the OBH, by making use of Eq.~(\ref{G_s11}) the explicit expressions for the diagonal $z$ component of the scattering part of the Green's tensor is expressed as
%
%%%%%%%%%%%%%%%%%%%%%%%%%%%%%%%%%%%%%%%%%%%%%%%%%%%%%%%%%%%%%%%%%
\begin{figure*}[t]
\begin{minipage}[b]{0.42\linewidth}
\centering
\includegraphics[width=\textwidth]{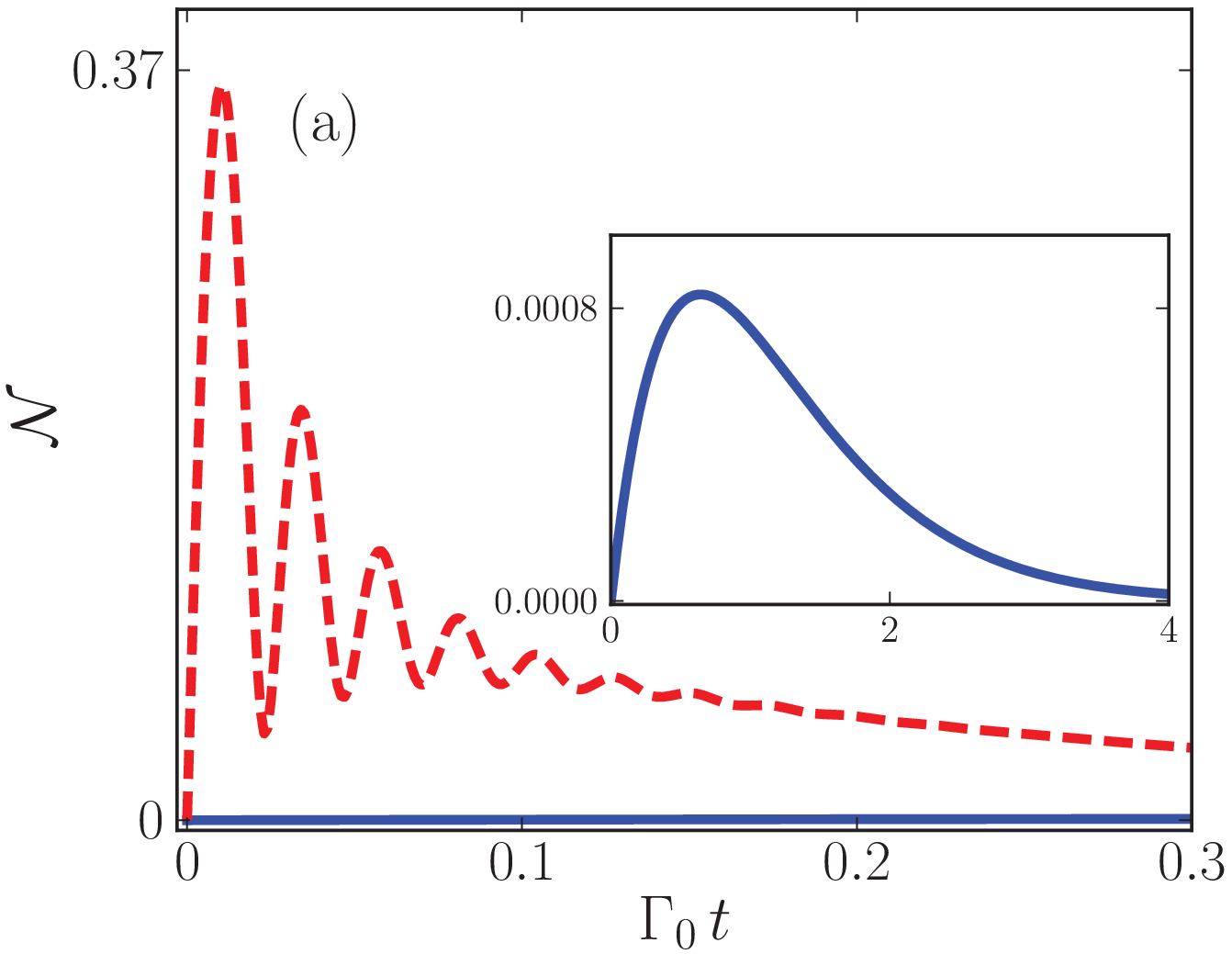}
\end{minipage}
\hspace{1cm}
\begin{minipage}[b]{0.42\linewidth}
\centering
\includegraphics[width=\textwidth]{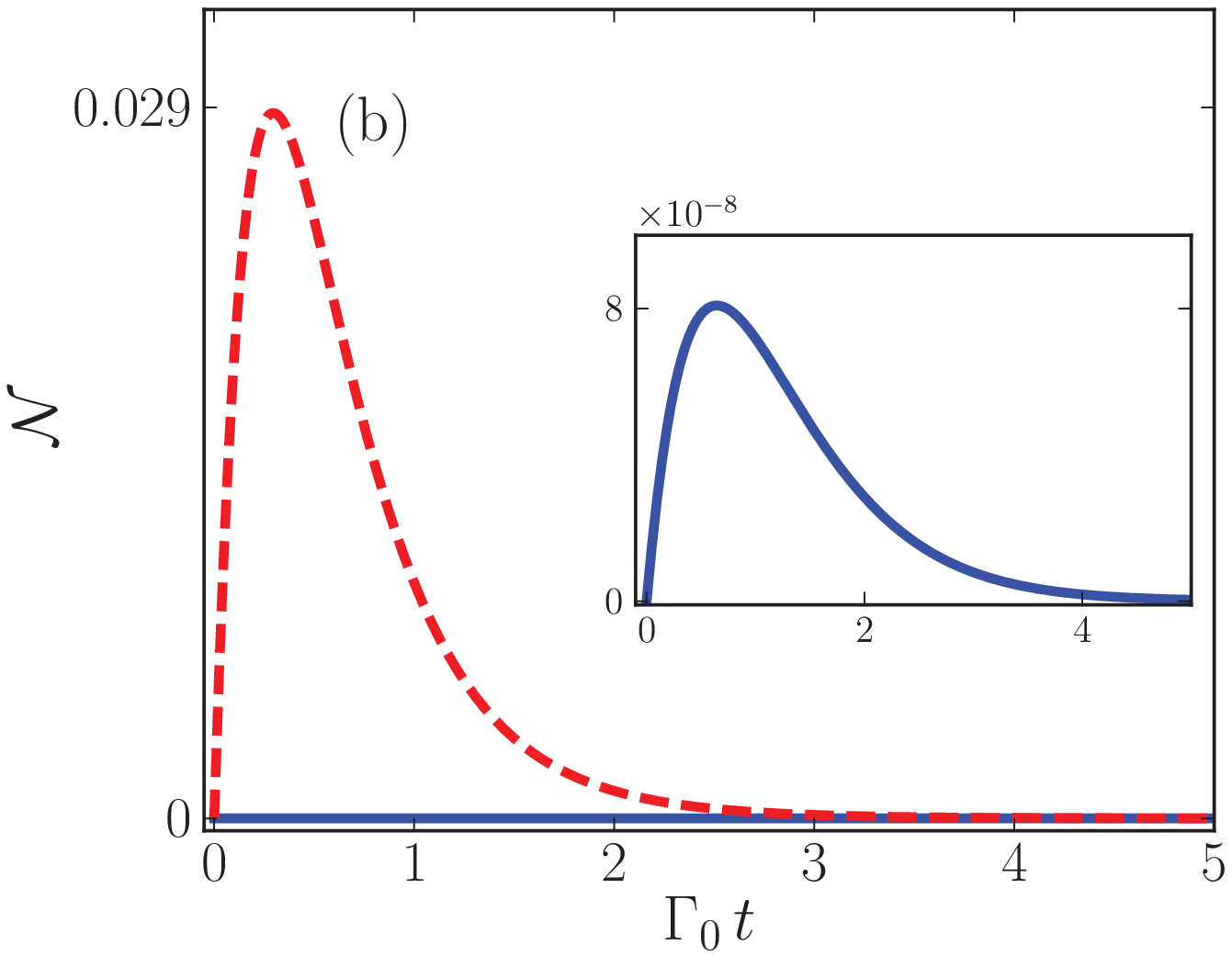}
\end{minipage}
\caption{The time evolution of the negativity ${\cal N}$ as a function of a dimensionless parameter  $ {\Gamma _0}{\rm{t}} $ for the case that the two atoms are placed  at position $ r=8.1\pi c/\omega_0$ outside and near the OBH (dashed red line), and in free space (solid blue line). The transition frequency of the two atoms is at non-resonance $\omega=0.1\omega_0$ (a), and resonance $\omega=\omega_0$ (b) with the resonance frequency of the OBH. The material absorbtion and dispersion of the OBH are described by the Lorentz model with parameters $ \omega_p= 0.1 \omega_0 $  and  $    {\gamma}= 0.01\omega_0 $. The insets show the magnified view of ${\cal N}$ versus $ {\Gamma _0}{\rm{t}} $ when the atoms are in free space.}
\label{Fig:2}
\end{figure*}
%%%%%%%%%%%%%%%%%%%%%%%%%%%%%%%%%%%%%%%%%%%%%%%%%%%%%%%%%%%%%%%%%
%
%\begin{eqnarray}\label{G_0zz}
%\bfsfG_{0,zz}({\rm{\bf{r}}},{\rm{\bf{r}}}')= ...,
%\end{eqnarray}
%
%and
%
\begin{eqnarray}\label{G_szz}
\bfsfG_{\rm S,zz}^{(11)}({\rm{\bf{r}}},{\rm{\bf{r}}}')& =& \frac{i}{{8\pi }}\int_{ - \infty }^\infty dh\sum_{n = 0}^\infty  {\frac{{(2 - \delta _n^0{\rm{)}}}}{{k _1^2}}}
 {\eta_1^2} C_{1V}^{11'} H_n^{(1)}({\eta _1}r) \nonumber \\
&&  \times H_n^{(1)}({\eta _1}r') \cos n(\varphi-\varphi') e^{i h(z-z')}.
\end{eqnarray}
By substituting the above equation into Eqs.~(\ref{Gamma ij}), $\Gamma_{\pm}$ in the presence of the OBH are given by
\begin{eqnarray}\label{Gamma+-}
\frac{{\Gamma^\pm}}{{{\Gamma _0}}} &= & \frac{3}{4}{\rm Re}\bigg[\int_{ - \infty }^\infty dh \sum_{n = 0}^\infty  {\frac{{(2 - \delta _n^0) }}{{k _1^3}}}
 {\eta_1^2}  H_n^{(1)}({\eta_1}r) \\
&&  \times \Big( J_n ({\eta _1}r)+C_{1V}^{11'} H_n^{(1)}({\eta_1}r)\Big) \big(1\pm(-1)^n \big)\bigg]. \nonumber
\end{eqnarray}
%
%
%\begin{eqnarray}\label{deltaAB}
%\frac{\delta_{AB}}{\Gamma _0}&=&\frac{3}{4}{\rm Im}\bigg[\int_{ - \infty }^\infty dh \sum_{n = 0}^\infty  {\frac{{(2 - \delta _n^0) }}{{k _1^3}}}
% {\eta_1^2}  (-1)^{n+1} H_n^{(1)}({\eta_1}r) \nonumber\\
%&&  \times \Big( J_n ({\eta _1}r)+C_{1V}^{11'} H_n^{(1)}({\eta_1}r)\Big) \bigg]. 
%\end{eqnarray}
%
where $ {{\rm{\Gamma }}_0} = \frac{{{{\tilde \omega }^3}_A{{d}_A}^2}}{{3\hbar \pi {\varepsilon _0}{c^3}}}$ is the free space decay rate and the unknown coefficient $ C^{11'}_{1V} $ is obtained by using the recurrence relation~(\ref{unknown coefficients C11}).
%
%Note that, in obtaining the Eq.~(\ref{deltaAB}), the Kramers-Kronig relation is applied to the Green's tensor of the system.
%
%%%%%%%%%%%%%%%%%%%%%%%%%%%%%%%%%%%%%%%%%%%%%%%%%%%%%%%%
\section{Entanglement  of the two two-level atoms}\label{Sec:Entanglement}
%%%%%%%%%%%%%%%%%%%%%%%%%%%%%%%%%%%%%%%%%%%%%%%%%%%%%%%%%%
%
We define the reduced density operator $\rho_a= {\rm Tr}_{field} | \psi  \rangle  \langle  \psi  |$, which is obtained
by tracing the density operator of the total system over the field degrees of freedom, to describe the atomic subsystem
in terms of the state vector~(\ref{time-dependent state vector of the system}) of the whole system.
%
%In the basis of the collective states, $| \pm  \rangle ,| 1 \rangle  =| l_A ,l_B \rangle ,| 4 \rangle =| u_A , u_B \rangle$,
%
In the absence of external driving fields, the two-atom
system is equivalent to a single four-level system composed of the ground state $| L \rangle  =| l_A ,l_B \rangle $, the upper state $| U \rangle =| u_A , u_B \rangle$, and two intermediate states $| + \rangle$ and $| -  \rangle$~\cite{Tanas2004,Dung2002}.
%The states $| + \rangle$ and $| -  \rangle$ decay to the two atom ground state $| l \rangle$	with the rates $\Gamma_+$ and $\Gamma_-$, respectively. The eigenenergies of $|\pm \rangle$ states are split by the dipole-dipole coupling.
%
In this basis, the reduced density operator $\rho_a$ is written as:
%Applying (15), we can find
%
\begin{eqnarray}\label{reduced density matrix}
\rho_a &=& | C_+|^2 | + \rangle \langle + |   + | C_-|^2 | - \rangle \langle - |\nonumber\\
&&     +C_+ C^*_-  | + \rangle \langle - |  +C_- C^*_+  | - \rangle \langle + |\nonumber\\
&&     +(1-{{\left| {{C}_{+}} \right|}^{2}}-{{\left| {{C}_{-}} \right|}^{2}}) | L \rangle \langle L | .
\end{eqnarray}
Now let us investigate the dynamics of entanglement between the two atoms. To characterize the quantum entanglement, there are various kinds of entanglement measures~\cite{Bennett1996,Wootters1998,Vidal2002}. We take
negativity as a well-known measure of mixed state entanglement for its simplicity as well as wide applicability. The entanglement negativity is defined by ${\cal N}= 1/2\sum_i (\left| {\mu _i} \right| - {\mu _i})$, where $ \mu_{i} $ are the eigenvalues of the partial transpose $\rho_a^{T_A}$. The negativity is $1$ for the maximally entangled states and $0$ for separable states.
For the reduced density matrix $\rho_a$, which describes a mixed state in the Hilbert space
${\cal H}_A \otimes {\cal H}_B$, its partial transposition with respect to the subsystem $A$ is
formally defined by taking the transpose of the matrix elements of $\rho_a$ with respect to the indices in subsystem $A$, i.e., $\rho_{a_{ ik,jl}}^{T_A}=\rho_{a_{ jk,il}}$ with $\rho_{a_{ ik,jl}}=\langle j,k| \rho |i,l\rangle$.
Applying~(\ref{reduced density matrix}), we can find
%
%\begin{widetext}
\begin{eqnarray}\label{negativity}
{\cal N}(t)&=& \bigg\{\Big(\frac{1}{4}({\rho_{a,++}} + {\rho_{a,--}})^2 +  \rho_{a,LL}^2 \\
&& - \exp [ - 2\Gamma t]
\cos^2(2{\delta_{AB}}t) \Big)^{1/2} - \rho_{a,LL}\bigg\}\bigg/{2},\nonumber
\end{eqnarray}
%\end{widetext}
where $\rho_{a,\pm\pm}$ and $\rho_{a,LL}$ are density matrix elements in the Dicke basis $| \pm \rangle$ and $| L \rangle$, respectively.
From this equation, the entanglement of the two atoms can be determined using the Eq.~(\ref{Gamma+-}), and Eq.~(\ref{deita ij}) together with the Kramers-Kronig relation, but the complexity of the resulting equation makes it difficult to predict the results analytically.

In Fig.~\ref{Fig:2}, the numerical results of the negativity~(\ref{negativity}) are plotted as a function of the dimensionless parameter  $  \Gamma _0 t $ for the non-resonance and resonance cases, namely, when the field frequency and the resonance frequency of the OBH satisfying the conditions $\omega=0.1\omega_0$ and $\omega=\omega_0$, respectively. For all our calculations, we assumed that $ r=8.1\pi c/\omega_0$.
%
%s well known that  is meaningless. In fact, two non-resonance and resonance situations for the two atoms in free space
The time evolution of ${\cal N}$ for the case that the atoms are placed in the free space and in the vicinity of the OBH are presented by the solid blue line and the dashed red, respectively. Recalling the Lorentzian factor $\varepsilon_L$ in Eq.~(\ref{material parameters of optical BH}), one might reasonably ask about the resonance frequency $\omega_0$ when the two atoms are in free space.
In fact, in this case, the involved parameters can become dimensionless by means of $\omega_A$. This means that we encounter with the atom-atom distance $ 16.2\pi c/\omega_A$ and $1.62\pi c/\omega_A$ for the resonance and non-resonance cases, respectively.

In Fig.~\ref{Fig:2} (a), it is seen that ${\cal N}$ is zero at the initial time $t = 0$. This is as it should be, since the initial state of the atomic subsystem is the product state $| u_A , l_B \rangle $. As time passes, the negativity shows an oscillatory behavior followed by a slow decay, which reflects that the two atoms are entangled due to the indirect interaction mediated through the OBH. The oscillatory behavior is observed at times shorter than $0.05\Gamma_0^{-1}$. This dynamic behavior can be easily understood from the dominant contribution of the oscillatory term with the frequency of the energy separation of the symmetric and antisymmetric states $2 \delta_{AB}$ in Eq.~(\ref{negativity}).
After the time  $ 0.05\Gamma _{0}^{-1}$, the time evolution of the population of the states $|\pm \rangle$ and $| L \rangle$ play an important role in decaying ${\cal N}$. Two states $|\pm \rangle$ are equally populated initially, but the decay of the state $\Gamma_+$ differs from $\Gamma_-$ in the presence of the OBH such that the symmetric state decay faster than the antisymmetric state (not shown here). Note that the decay value $\Gamma_+$ and $\Gamma_-$ are also different from each other for the two atoms in free space~\cite{Tanas2004}. However, this difference that occurs by mediation of the OBH is much larger.
For long times, the symmetric and antisymmetric states are depopulated, consequently, the terms $\rho_{a,\pm\pm}$ and $\rho_{a,LL}$ in~(\ref{negativity}) decay to zero and unit, respectively, and the negativity goes to zero.

In Fig.~\ref{Fig:2} (b), the time evolution of ${\cal N}$ is depicted as a function of a dimensionless parameter ${{\Gamma }_{0}}\,t$ for the
case that the two atoms are at resonance with the OBH $\omega =\omega_0$.
The negativity is characterized by a fast initial increase and slow decrease after reaching maximum.
In the presence of the OBH, the maximum value of the negativity is $0.029$ (dashed red line), which is one magnitude lower than the value for the non-resonant case.
With the high loss that occurs at resonance frequency, the OBH absorbs the spontaneously emitted photon before it exits the shell. One thus expect that
a lossy OBH decrease the amount of entanglement created between two atoms.
It is also seen that there is no oscillations at short times as a result of the significant reduction of the dipole-dipole coupling ${{\delta }_{AB}}$ at resonance frequency. This is because, from Eq.~(\ref{deita ij}), the behavior of $\delta_{AB} $ depends on
the material environment and, as well as the distance of the two atoms via the Green's tensor.
Unlike to this situation, due to large distance of the atoms from each other the negativity for the case that the two atoms are in free space experiences a sharp decrease from $8\times 10^{-4}$ at non-resonance case to $8\times 10^{-8}$  at resonant case.
We eventually find that the entanglement can be created between two atoms in vicinity of the OBH, and is well in agreement with the results of the schwarzschild BH presented in~\cite{Hu2011}.
\section{Summery and conclusion}\label{Sec:conclusion}
In this paper, we study a macroscopic system consisting of two two-level atoms that weakly interact with the electromagnetic field prepared in its vacuum state
in vicinity of the OBH. We consider the case when one atom was in the excited state and the other in the ground state.
For the given system, based on a canonical quantization scheme presented for the electromagnetic field interacting
with atomic systems in the presence of absorptive and dispersive dielectric media, we derive the reduced density operator of the
atomic system and, investigate the collective behavior of the atoms in Markovian approximation.
As the formalism shows, these expressions have been expressed in terms of the Green's tensor of the system.
We have modeled the artificial black hole by a multilayered cylindrical structure with homogeneous material parameters in a stepwise manner. In this way, we used the Green's tensor of a multilayer cylinder structure and, hereby, numerical calculations are performed for the negativity as a measure of entanglement.
The time evolution of the negativity between the atoms has been discussed for two non-resonance and resonance cases.
%the time evolution of entanglement of the atoms in free space and near the optical BH have been investigated. photon exchange process may be mediated by the 
The results show that the photon exchange process is mediated by the OBH can induce entanglement between the atoms, of course, if the photon is not complectly absorbed by the shell.
\appendix
%%%%%%%%%%%%%%%%%%%%%%%%%%%%%%%%%%%%%%%%%%%%%%%%%%%%%%%%%%%%%%%%%%%%%%%
%
\section{Green tensor of the system}\label{App:Green tensor}
The calculation of the electromagnetic Green's tensor of a multilayer cylinder structure has been extracted previously in~\cite{Li2000}.
Based on the method of scattering superposition and the linearity of the Helmholtz equation~(\ref{Helmholtz equation}), the Green's tensor of the system  can be decomposed into a vacuum component and a scattering contribution
\begin{eqnarray}\label{3b}
\bfsfG_e^{fs} = {\bfsfG_{0}}({\bf{r}},{\bf{r'}})\delta_f^s + \bfsfG_{\rm S}^{(fs)}({\bf{r}},{\bf{r'}}),
\end{eqnarray}
where $f$ and $s$ indicate the regions of the field and source points, respectively. The Green tensor $\bfsfG_{0}$ describes the propagation inside in free space, whereas the scattering contribution requires knowledge about the reflection and
transmission coefficients at the cylindrical interfaces. The form of these two contributions in the cylindrical coordinate system are given by:
\begin{eqnarray}\label{vacuum component of Green}
&&{{\rm{\bf G}}_0}({\rm{\bf r}},{\rm{\bf r}}') = \frac{ -{{{\hat {\bf r}\hat{\bf r} }}\delta ({\rm{\bf r}} - {\rm{\bf r}}')}}{{k_s^2}} + \frac{i}{{8\pi }}\int_{ - \infty }^\infty  {dh\sum_{n = 0}^\infty  {\frac{{(2 - \delta _n^0)}}{{\eta _s^2}}} } \\
&&   \left\{ {\begin{array}{*{20}{c}}
{{{\bf M}}_{_o^en{\eta _s}}^{(1)}(h){{{{\bf M'}}}_{_o^en{\eta _s}}}( - h) + {{\bf N}}_{_o^en{\eta _s}}^{(1)} (h){{{{\bf N'}}}_{_o^en{\eta _s}}}( - h)}\;\; r>r'\\
{{{\rm{\bf M}}_{_o^en{\eta _s}}}(h){\bf M'}^{(1)}_{_o^e n{\eta _s}}( - h) + {{{\bf N}}_{_o^en{\eta _s}}} (h){\bf N'}^{(1)}_{_o^en{\eta _s}}( - h)}\;\;r<r'
\end{array}} \right.\nonumber
\end{eqnarray}
and
\begin{eqnarray}\label{scattering contribution of Green}
&&\bfsfG_{\rm S}^{(fs)}({\rm{\bf{r}}},{\rm{\bf{r}}}') =\frac{i}{{8\pi }}\int_{ - \infty }^\infty  {dh\sum_{n = 0}^\infty  {\frac{{(2 - \delta _n^0)}}{{\eta _s^2}}} } \nonumber\\
&& \times \bigg\{ {(1 - \delta _f^N)} {\rm{\bf{M}}}_{_o^en{\eta _f}}^{(1)}(h)\left[ {(1 - \delta _s^1)C_{1H}^{fs}{\rm{\bf{M}}}_{_o^en{\eta _s}}^\prime ( - h)} \right.\nonumber\\
&& + \left. {(1 - \delta _s^N)C_{1H}^{fs'}{\rm{\bf{M}}}_{_o^en{\eta _s}}^{\prime (1)}( - h)} \right]\nonumber\\
&& + (1 - \delta _f^N){\rm{\bf{N}}}_{_o^en{\eta _f}}^{(1)}(h)\left[ {(1 - \delta _s^1)C_{1V}^{fs}{\rm{\bf{N}}}_{_o^en{\eta _s}}^\prime ( - h)} \right.\nonumber\\
&& + (1 - \delta _s^N)\left. {C_{1V}^{fs'}{\rm{\bf{N}}}_{_o^en{\eta _s}}^{\prime (1)}( - h)} \right]\nonumber\\
&& + (1 - \delta _f^N){\rm{\bf{N}}}_{_e^on{\eta _f}}^{(1)}(h)\left[ {(1 - \delta _s^1)C_{2H}^{fs}{\rm{\bf{M}}}_{_o^en{\eta _s}}^\prime ( - h)} \right.\nonumber\\
&& + (1 - \delta _s^N)\left. {C_{2H}^{fs'}{\rm{\bf{M}}}_{_o^en{\eta _s}}^{\prime (1)}( - h)} \right]\nonumber\\
&& + (1 - \delta _f^N){\rm{\bf{M}}}_{_e^on{\eta _f}}^{(1)}(h)\left[ {(1 - \delta _s^1)C_{2V}^{fs}{\rm{\bf{N}}}_{_o^en{\eta _s}}^\prime ( - h)} \right.\nonumber\\
&& + (1 - \delta _s^N)\left. {C_{2V}^{fs'}{\rm{\bf{N}}}_{_o^en{\eta _s}}^{\prime (1)}( - h)} \right]\nonumber\\
&& + (1 - \delta _f^1){{\rm{\bf{M}}}_{_o^en{\eta _f}}}(h)\left[ {(1 - \delta _s^1)C_{3H}^{fs}{\rm{\bf{M}}}_{_o^en{\eta _s}}^\prime ( - h)} \right.\nonumber\\
&& + (1 - \delta _s^N)\left. {C_{3H}^{fs'}{\rm{\bf{M}}}_{_o^en{\eta _s}}^{\prime (1)}( - h)} \right]\nonumber\\
&& + (1 - \delta _f^1){{\rm{\bf{N}}}_{_o^en{\eta _f}}}(h)\left[ {(1 - \delta _s^1)C_{3V}^{fs}{\rm{\bf{N}}}_{_o^en{\eta _s}}^\prime ( - h)} \right.\nonumber\\
&& + (1 - \delta _s^N)\left. {C_{3V}^{fs'}{\rm{\bf{N}}}_{_o^en{\eta _s}}^{\prime (1)}( - h)} \right]\nonumber\\
&& + (1 - \delta _f^1){{\rm{\bf{N}}}_{_{_e^o}n{\eta _f}}}(h)\left[ {(1 - \delta _s^1)C_{4H}^{fs}{\rm{\bf{M}}}_{_o^en{\eta _s}}^\prime ( - h)} \right.\nonumber\\
&& + (1 - \delta _s^N)\left. {C_{4H}^{fs'}{\rm{\bf{M}}}_{_o^en{\eta _s}}^{\prime (1)}( - h)} \right]\nonumber\\
&& + (1 - \delta _f^1){{\rm{\bf{M}}}_{_{_e^o}n{\eta _f}}}(h)\left[ {(1 - \delta _s^1)C_{4V}^{fs}{\rm{\bf{N}}}_{_o^en{\eta _s}}^\prime ( - h)} \right.\nonumber\\
&& + (1 - \delta _s^N) {\left. {C_{4V}^{fs'}{\rm{\bf{N}}}_{_o^en{\eta _s}}^{\prime (1)}( - h)} \right]} \bigg\},
\end{eqnarray}
where the prime denotes the coordinates $ (r',\phi ',z') $ of the source, and the eigenvalue, ${\eta_f}$, and the propagating constant, $k_f$, in the $f$th layer satisfy the relation $h^2=k_f^2-\eta_f^2$. Here, ${{\mathop{\rm \bf{M}}\nolimits} _{_o^en{\eta_f}}}(h)$ and ${{\mathop{\rm \bf{N}}\nolimits} _{_o^en{\eta _f}}}(h)$ are the cylindrical wave vector functions, and defined as~\cite{Tai1994}
\begin{subequations}\label{71b}
\begin{eqnarray}
{{\mathop{\rm \bf{M}}\nolimits} _{_o^en{\eta _f}}}(h) &= &\bigg[  \mp \frac{{n{Z_n}({\eta _f}r)}}{r}\begin{array}{*{20}{c}}
{{\sin}}\\
{{\cos}}
\end{array}(n\phi )\hat{\bf r} \\
&&   - \frac{{d{Z_n}({\eta _f}r)}}{{dr}}\begin{array}{*{20}{c}}
{{\cos}}\\
{{\sin}}
\end{array}(n\phi )\hat {\boldsymbol \phi}  \bigg] {\exp}(ihz),\nonumber\label{71ab}\\
{{\mathop{\rm\bf{N} }\nolimits}_{_o^en{\eta _f}}}(h)& =& \frac{1}{{{k_f}}}\bigg[ ih\frac{{d{Z_n}({\eta _f}r)}}{{dr}} \begin{array}{*{20}{c}}
{{\rm{cos}}}\\
{{\rm{sin}}}
\end{array}(n\phi )\hat{\bf r} \nonumber\\
&& \mp \frac{{ihn}}{r}{Z_n}({\eta_f}r) \begin{array}{*{20}{c}}
{{\sin}}\\
{{\cos}}
\end{array}(n\phi )\hat{\boldsymbol \phi} \\ 
&&+ {\eta ^2}{Z_n}({\eta _f}r) \begin{array}{*{20}{c}}
{{\cos}}\\
{{\sin}}
\end{array}(n\phi )\hat{\bf z}  \bigg]{\exp}(ihz).\nonumber\label{71bb}
\end{eqnarray}
\end{subequations}
Here, $Z_{n}({\eta _f}r)$ and $Z_{n}^{(1)}({\eta _f}r)$ are, respectively, the first-type cylindrical Bessel function $J_n({\eta _f}r)$ and the third-type cylindrical Bessel function or the first-type cylindrical Hankel function $H_n^{(1)}({\eta _f}r)$. 

Due to the fact that in our case the two-atom system is in free space outside the OBH, both the observation
point ${\bf r}$ and the source point ${\bf r}'$ are located in first layer outside the shell. In the following, we therefore set $ s=f=1 $ in Eq.~(\ref{scattering contribution of Green}) and arrive at
%
%\begin{eqnarray}\label{G_011}
%&&\bfsfG_{0}^{(11)}({\rm{\bf{r}}},{\rm{\bf{r}}}') =  - \frac{{{\rm{\hat {\bf r}\hat{\bf r} }}\delta ({\rm{\bf r}} - {\rm{\bf r}}')}}{{k_1^2}} + \frac{i}{{8\pi %}}\int\limits_{ - \infty }^\infty  {dh\sum\limits_{n = 0}^\infty  {\frac{{(2 - \delta _n^0)}}{{\eta _1^2}}} } \\
%&& \times \left\{ {\begin{array}{*{20}{c}}
%{{\rm{\bf M}}_{_o^en{\eta _1}}^{(1)}(h){{{\rm{\bf M'}}}_{_o^en{\eta _1}}}( - h) + {\rm{\bf N}}_{_o^en{\eta _1}}^{(1)}(h){{{\rm{\bf N'}}}_{_o^en{\eta _1}}}( - %h)}\;\;\;\;\;\; r>r'\\
%{{{\rm{\bf M}}_{_o^en{\eta _1}}}(h){{{\rm{\bf M'}}}^{(1)}}_{_o^en{\eta _1}}( - h) + {{\rm{\bf N}}_{_o^en{\eta _1}}}(h){{{\rm{\bf N'}}}^{(1)}}_{_o^en{\eta _1}}( - %h)}\;\;\;\;\;\;r<r'
%\end{array}} \right.\;
%\end{eqnarray}
%
%and
%
\begin{eqnarray}\label{G_s11}
\bfsfG_{\rm S}^{(11)}(\bf r,\bf r') &=& \frac{i}{{8\pi }}\int\limits_{ - \infty }^\infty  {dh\sum\limits_{n = 0}^\infty  {\frac{{(2 - \delta _n^0)}}{{\eta _1^2}}} } \nonumber \\ 
&\times & \left[ {C_{1H}^{11'}} \right.{\rm{\bf M}}_{_o^en{\eta _1}}^{(1)}(h){\rm{\bf M'}}_{_o^en{\eta _1}}^{(1)}( - h) \nonumber\\ 
&+& C_{1V}^{11'}{\rm{\bf N}}_{_o^en{\eta _1}}^{(1)}(h){\rm{\bf N'}}_{_o^en{\eta _1}}^{(1)}( - h)\, \\ 
&+& C_{2H}^{11'}{\rm{\bf N}}_{_e^on{\eta _1}}^{(1)}(h){\rm{\bf M'}}_{_o^en{\eta _1}}^{(1)}( - h) \nonumber\\ 
&+&\left. { C_{2V}^{11'}{\rm{\bf M}}_{_e^on{\eta _1}}^{(1)}(h){\rm{\bf N'}}_{_o^en{\eta _1}}^{(1)}( - h)} \right].\nonumber
\end{eqnarray}
To obtain the unknown coefficients in equation above, we introduce the following recurrence relation:
\begin{eqnarray}\label{recurrence relation}
{\rm T}_f^{(H,V)} = {\left[ {R_{f(ij)}^{(H,V)}} \right]_{4 \times 4}} = {\left[ {F_{(f + 1)f}^{(H,V)}} \right]^{ - 1}}.\left[ {F_{ff}^{(H,V)}} \right],\,\,\,
\end{eqnarray}
where
\begin{eqnarray}\label{T_HV}
{\rm T}_{(H,V)}^{(K)} &=& {\left[ {T_{ij}^{K(H,V)}} \right]_{4 \times 4}}\\
&=& \left[ {{\rm T}_{N - 1}^{(H,V)}} \right]\left[ {{\rm T}_{N - 2}^{(H,V)}} \right] \cdots \left[ {{\rm T}_{K + 1}^{(H,V)}} \right]\left[ {{\rm T}_K^{(H,V)}} \right],\nonumber
\end{eqnarray}
with $ {\left[ {F_{(f + 1)f}^{(H,V)}} \right]^{ - 1}} $  is the inverse matrix of  $  {F_{(f+1)f}^{(H,V)}} $.
The transmission matrices $F^{H,V}$ for TE and TM waves are, respectively, defined as:
\begin{widetext}
\begin{subequations}\label{transmission matrices}
\begin{eqnarray}
F_{jm}^H = \left[ {\begin{array}{*{20}{c}}
{\frac{{\partial \left[ {H_n^{(1)}({\eta _j}{a_m})} \right]}}{{\partial {a_m}}}}&{ \mp \frac{{{\zeta _j}H_n^{(1)}({\eta _j}{a_m})}}{{{a_m}}}}&{\frac{{\partial \left[ {{J_n}({\eta _j}{a_m})} \right]}}{{\partial {a_m}}}}&{ \mp \frac{{{\zeta _j}{J_n}({\eta _j}{a_m})}}{{{a_m}}}}\\
0&{{\ell _j}H_n^{(1)}({\eta _j}{a_m})}&0&{{\ell _j}{J_n}({\eta _j}{a_m})}\\
{ \pm \frac{{{\zeta _j}{\tau _j}H_n^{(1)}({\eta _j}{a_m})}}{{{a_m}}}}&{\frac{{{\tau _j}\partial \left[ {H_n^{(1)}({\eta _j}{a_m})} \right]}}{{\partial {a_m}}}}&{ \pm \frac{{{\zeta _j}{\tau _j}{J_n}({\eta _j}{a_m})}}{{{a_m}}}}&{\frac{{{\tau _j}\partial \left[ {{J_n}({\eta _j}{a_m})} \right]}}{{\partial {a_m}}}}\\
{{\tau _j}{\ell _j}H_n^{(1)}({\eta _j}{a_m})}&0&{{\tau _j}{\ell _j}{J_n}({\eta _j}{a_m})}&0
\end{array}} \right],\\
\nonumber\\
\begin{array}{l}
F_{jm}^V = \left[ {\begin{array}{*{20}{c}}
{ \pm \frac{{{\zeta _j}H_n^{(1)}({\eta _j}{a_m})}}{{{a_m}}}}&{\frac{{\partial \left[ {H_n^{(1)}({\eta _j}{a_m})} \right]}}{{\partial {a_m}}}}&{ \pm \frac{{{\zeta _j}{J_n}({\eta _j}{a_m})}}{{{a_m}}}}&{\frac{{\partial \left[ {{J_n}({\eta _j}{a_m})} \right]}}{{\partial {a_m}}}}\\
{{\ell _j}H_n^{(1)}({\eta _j}{a_m})}&0&{{\ell _j}{J_n}({\eta _j}{a_m})}&0\\
{\frac{{{\tau _j}\partial \left[ {H_n^{(1)}({\eta _j}{a_m})} \right]}}{{\partial {a_m}}}}&{ \mp \frac{{{\zeta _j}{\tau _j}H_n^{(1)}({\eta _j}{a_m})}}{{{a_m}}}}&{\frac{{{\tau _j}\partial \left[ {{J_n}({\eta _j}{a_m})} \right]}}{{\partial {a_m}}}}&{ \mp \frac{{{\zeta _j}{\tau _j}{J_n}({\eta _j}{a_m})}}{{{a_m}}}}\\
0&{{\tau _j}{\ell _j}H_n^{(1)}({\eta _j}{a_m})}&0&{{\tau _j}{\ell _j}{J_n}({\eta _j}{a_m})}
\end{array}} \right]\label{15b}.\\
\end{array}                  
\end{eqnarray}
\end{subequations}
\end{widetext}
where $ a_{m} \, (m=1,2,...,N-1)$ denote the radius of different layers. Here, the following parameters have been introduced to
simplify the symbolic calculations
\begin{eqnarray}\label{symbolic definition}
{\tau _j} = \sqrt {\frac{{{\varepsilon _j}}}{{{\mu _j}}}} ,\;\;\;   \quad {\zeta _j} = \frac{{ihn}}{{{k_j}}},\;\; \;  \quad {\ell _j} = \frac{{{{({\eta _j})}^2}}}{{{k_j}}}.
\end{eqnarray}
Now by using Eqs.(\ref{recurrence relation})-(\ref{symbolic definition}), the unknown coefficients entered to our calculations in Eq.~(\ref{G_szz}) are obtained as
\begin{eqnarray}\label{unknown coefficients C11}
C_{1(H,V)}^{11'} = \frac{{T_{12}^{1(H,V)}T_{23}^{1(H,V)} - T_{22}^{1(H,V)}T_{13}^{1(H,V)}}}{{T_{11}^{1(H,V)}T_{22}^{1(H,V)} - T_{12}^{1(H,V)}T_{21}^{1(H,V)}}}.\,\,\,\,
\end{eqnarray}
%For further details, the interested reader to Ref.~\cite{Li2000}.
%

%%%%%%%%%%%%%%%%%%%%%%%%%%%%%%%%%%%%%%%%%%%%%%%%%%%%%%%%%%%%%%%%%%%%%%%%%

\end{document}